\documentclass[preprint]{revtex4-2}
\usepackage[utf8]{inputenc}
\usepackage[T1]{fontenc}
\usepackage{graphicx}
\usepackage{amsmath}
\usepackage{amssymb}
\usepackage[]{bibunits}
\defaultbibliographystyle{apsrev4-1} 
\usepackage{threeparttable}
\defaultbibliography{references}
\begin{document}

\title{Implantable silicon neural probes with nanophotonic phased arrays for single-lobe beam steering}

\author{Fu-Der Chen\textsuperscript{1,2,3,*}}
\author{Ankita Sharma\textsuperscript{1,2,3,*,$\dagger$}}
\author{Tianyuan Xue\textsuperscript{1,2}}
\author{Youngho Jung\textsuperscript{1} }
\author{Alperen Govdeli\textsuperscript{1,2}}
\author{Jason C. C. Mak\textsuperscript{2} }
\author{Homeira Moradi Chameh\textsuperscript{4}}
\author{Mandana Movahed\textsuperscript{4}}
\author{Michael G.K. Brunk\textsuperscript{1,3}}
\author{Xianshu Luo\textsuperscript{5}}
\author{Hongyao Chua\textsuperscript{5}}
\author{Patrick Guo-Qiang Lo\textsuperscript{5}}
\author{Taufik A Valiante\textsuperscript{2,3,4,6,7}}
\author{Wesley D. Sacher\textsuperscript{1,3}}
\author{Joyce K.S. Poon\textsuperscript{1,2,3,$\dagger$}}

\affiliation{\textsuperscript{1}Max Planck Institute of Microstructure Physics, Weinberg 2, 06120 Halle, Germany}

\affiliation{\textsuperscript{2}Department of Electrical and Computer Engineering, University of Toronto, 10 King’s College Road, Toronto, Ontario M5S 3G4, Canada}

\affiliation{\textsuperscript{3} Max Planck-University of Toronto Centre for Neural Science and Technology}

\affiliation{\textsuperscript{4}Krembil Brain Institute, University Health Network, Toronto, Ontario, Canada}

\affiliation{\textsuperscript{5}Advanced Micro Foundry Pte. Ltd., 11 Science Park Road, Singapore Science Park II, 117685, Singapore}

\affiliation{\textsuperscript{6}Division of Neurosurgery, Department of Surgery, Toronto Western Hospital, University of Toronto, Toronto, Ontario, Canada}

\affiliation{\textsuperscript{7}Institute of Biomedical Engineering, University of Toronto, Toronto, Ontario, Canada}

\affiliation{\textsuperscript{*}These authors contributed equally to this work}

\affiliation{\textsuperscript{$\dagger$}Corresponding authors: ank.sharma@mail.utoronto.ca, joyce.poon@mpi-halle.mpg.de}

\begin{abstract}

In brain activity mapping experiments using optogenetics, patterned illumination is crucial for deterministic and localized stimulation of neurons. However, due to optical scattering in brain tissue, light-emitting implantable devices are needed to bring precise patterned illumination to deep brain regions. A promising solution is silicon neural probes with integrated nanophotonic circuits that form tailored beam emission patterns without lenses. Here, we demonstrate neural probes with grating-based light emitters that generate a single steerable light beam across $> 60\%$ of the steering range with $\ge 4$ dB of background suppression for optogenetic photostimulation. The light emitters, optimized for blue or amber light, combine end-fire optical phased arrays with slab gratings to suppress higher-order sidelobes. \emph{In vivo} optogenetic photostimulation was demonstrated in Thy1-ChR2-YFP mice. These neural probes, capable of continuous single beam scanning, offer a promising avenue for optogenetic stimulation with cellular-level spatial precision in the deep brain.

\end{abstract}

\begin{bibunit}
\flushbottom
\maketitle
\thispagestyle{empty}

\section{Introduction}

Optogenetics with patterned photostimulation combines spatially targeted optical actuation of neurons with cell-type specificity to enable the investigation of synaptic connectivity in neuronal circuits at cellular resolutions. Discrete optical beam scanners, such as galvanometers and acousto-optic deflectors \cite{Wang2007, Petreanu2009, Resta2022}, spatial light modulators \cite{Forli2018,Marshel2019}, and digital micromirror devices \cite{Kissinger2020} are commonly used to impart patterns on an optical beam for spatially targeted photostimulation. However, these components form illumination patterns in free space, and the penetration of visible to near-infrared light (400 - 1100 nm) is limited by attenuation to only about 1 mm from the brain surface in rodents \cite{Abaya2012}. To bring light into deep brain regions, implantable optical devices are being investigated, including optical fibers \cite{Aravanis2007,Szabo2014,Pisanello2017}, miniature gradient index (GRIN) lenses \cite{Stamatakis2018,Accanto2019,Jennings2019,Zhang2022}, and silicon (Si) neural probes with micro light-emitting diodes (\textmu LEDs) \cite{Wu2015,Scharf2016,Kim2020,Voroslakos2022}, organic LEDs \cite{Taal2023}, or integrated nanophotonic waveguides \cite{Segev2016,Libbrecht2018,Mohanty2020,Lanzio2021, Sacher2019,Neutens2023}.

Among these options, nanophotonic waveguide-based Si neural probes are promising for delivering light with high precision. A rich variety of light emission patterns can be achieved using waveguide-based grating emitters on the implant without the need for any lenses or discrete optics. In contrast, light-emitting implants, such as single-core optical fibers without wavefront compensation and LED probes, emit diffracting Gaussian and Lambertian beam profiles, respectively. Furthermore, nanophotonic waveguide probes have small form factors with widths and thicknesses $\leq$ 100 {\textmu}m and sharp tips, which ease surgical implantation and displace less tissue compared to fiber bundles and GRIN lenses, which have typical diameters exceeding 300 {\textmu}m. Through the design of grating emitters, we have demonstrated a wide range of beam patterns for optogenetic applications in the visible spectrum, including low-divergence beams, light sheets, focused, and steerable beams \cite{Sacher2019, Sacher2021, Sacher2022, xue2024implantable, Chen2023}.

Steerable beams provide dynamically patterned illumination by serially scanning a beam across a continuous region in tissue samples. This approach is valuable for localized spatial mapping of neuronal connections within a circuit \cite{Petreanu2009, Wang2007}. Optical phased arrays (OPAs) are common photonic circuits to realize steerable optical beams without mechanically moving parts \cite{raval_integrated_2018, Shin2020, sun_parallel_2021,  Sacher2022, Notaros2023}. Emitting a narrow beam with minimal divergence from an OPA requires a large aperture \cite{Baets2009}, which is not compatible with implantable devices due to size constraints. However, the resolution of optogenetic stimulation is generally limited to the size of the cell body partly because of the low conductance of individual opsin proteins to activate action potentials \cite{Lees2024}. Furthermore, microelectrodes on a neural probe have detection ranges up to about 140 \textmu m \cite{Henze2000}. Therefore, light beams with widths matching the diameter of individual cells (10-20 \textmu m) over a propagation distance of $\lesssim 140$ \textmu m already offer sufficient spatial precision. The relaxed criterion on the spatial resolution opens the opportunity to realize highly compact OPAs on neural probes to bring steerable light beams into deep brain regions \cite{Sacher2022}. Although \cite{Sacher2022} showed passive OPAs emitting beams with widths that matched cell diameters ($< 23$ {\textmu}m full width at half maximum (FWHM) beam width at a distance of 50 to 150 {\textmu}m from the probe), a major shortcoming was the emission of multiple beams (i.e., grating orders). The grating pitch in \cite{Sacher2022} exceeded the $\lambda/2$ criterion for single-beam emission in OPAs \cite{Shin2020}, and it could not be reduced due to fabrication limitations and inter-waveguide crosstalk. It is possible to emit a single steerable beam at visible wavelengths without satisfying the $\lambda/2$ pitch criterion, such as by using an aperiodic pitch \cite{Shin2020} or microcantilevers \cite{Sharif2023}. However, these approaches are not suited to implantable Si neural probes because of their size, thermally tuned phase shifters, or mechanical actuators, which can damage the brain tissue.

Here, we overcome the challenge of scanning a single-lobe beam at depth in brain tissue with passive OPAs that combine end-fire phased arrays with slab gratings on Si neural probes. The slab grating design consists of an optional free-propagation region (FPR) for beam formation in the waveguide followed by a grating to radiate the beam out of the plane. This OPA circuit achieves single-beam emission due to (1) an in-plane emitter pitch of about $\lambda$, made possible by the short waveguides in the end-fire phased array, and (2) the slab grating that cuts off higher-angle sidelobes. We study the steering ranges in slab grating designs with concave, straight, and convex curvatures that operate in the blue or amber wavelength ranges. To demonstrate that these novel OPAs emit sufficient optical power, probes with sidelobe-free OPAs and integrated titanium nitride (TiN) electrodes were used for photostimulation and recording in \emph{in vivo} awake head-fixed animal experiments. These compact single-lobe beam steering OPAs in the visible spectrum on implantable neural probes emit dynamic patterned illumination over a continuous region, a key step toward precise spatial mapping of neuronal activity in deep brain regions.

\section{Results}

\subsection{OPA Neural Probes on 200-mm Silicon Wafers}

 \begin{figure}[!ht]
\centering
\includegraphics[width=1\linewidth]{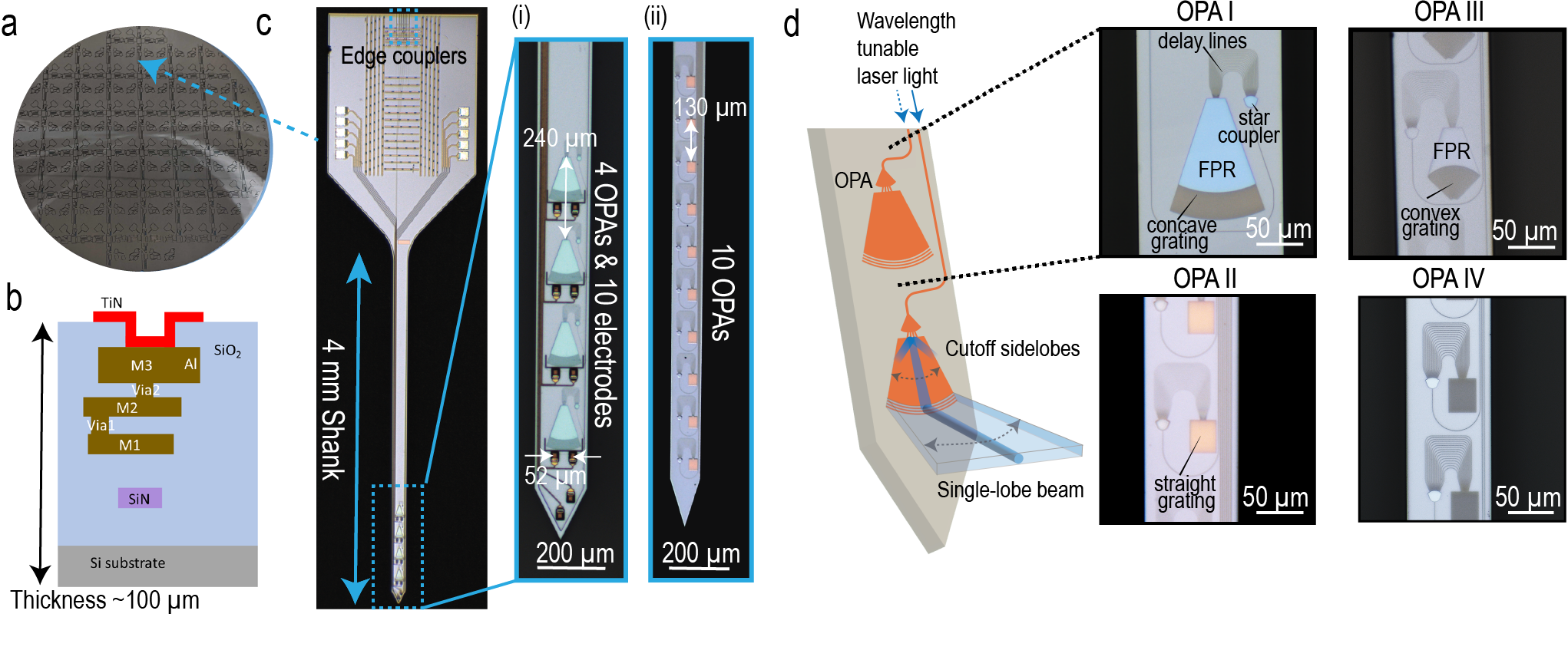}
\caption{\textbf{a} 200-mm diameter Si wafers containing silicon neural probes. \textbf{b} Cross-sectional view of the wafer in \textbf{a}. \textbf{c} Photograph of a beam steering neural probe with inset (i) showing an optical micrograph of the 154 \textmu m-wide shank containing 4 OPA emitters and 10 microelectrodes. The pitch between the OPAs is 240 \textmu m and the horizontal pitch between the electrodes is 52 \textmu m. Inset (ii) optical micrograph of a photonics-only shank variant containing 10 OPA emitters at a pitch of 130 \textmu m on a shank with a width of 100 {\textmu}m. The photonic devices span roughly 1.4 mm along the shank. \textbf{d } Illustration of the single-lobe OPA operation and annotated micrographs of four OPA designs with three of them having different slab grating curvatures. OPA Type IV shares a similar design to OPA Type II but is optimized for amber wavelengths. For all OPA designs, an end-fire array coupled to a slab grating enables the emission of a single steerable beam.}
\label{fig:device}
\end{figure}

The OPA neural probes were fabricated on 200-mm diameter Si wafers (see Fig.\ref{fig:device}a) at Advanced Micro Foundry (AMF) using the process reported in \cite{Chen2023}. A cross-sectional view of these wafers is depicted in Fig.\ref{fig:device}b. Waveguides were patterned in silicon nitride (SiN) formed by plasma-enhanced chemical vapour deposition (PECVD) or low pressure chemical vapour deposition (LPCVD). To facilitate electrophysiological recordings, the neural probe platform included biocompatible titanium nitride (TiN) deposited on the surface for microelectrodes and three aluminum metal layers for routing \cite{Chen2023}. 
 
A beam-steering neural probe is shown in Fig.\ref{fig:device}c. Input laser light was coupled from a custom multicore fiber to the chip using edge couplers \cite{Sacher2019}. The probes operated in the blue (450--490 nm) or amber (575--600 nm) part of the spectrum. The nanophotonic circuits for blue-light emitting probes were defined in a 150 nm thick PECVD SiN waveguide layer (refractive index between 1.82 and 1.9 at 488 nm) or 120 nm thick LPCVD SiN waveguide layer (refractive index of 2.04 at 488 nm). Neural probes for the amber wavelengths used a 200 nm thick LPCVD SiN layer. Figure \ref{fig:device}c inset (i) shows a hybrid neural probe design, which integrated four OPAs at a pitch of 240 \textmu m and ten 20 $\times$ 20 {\textmu}m$^{2}$ TiN electrodes on a 154 \textmu m-wide shank and (ii) shows a photonic-only probe design, which had ten OPAs at a pitch of 130 \textmu m spanning approximately 1.4 mm along a 100 \textmu m-wide shank.

We investigated four different sidelobe-free OPA designs integrated onto implantable neural probes. The specifications of these designs are summarized in Table \ref{table:design}. The OPA designs were passive to minimize the risk of tissue heating. Each OPA consisted of a star coupler, which divided the light from an input waveguide into 16 delay line waveguides. The delay lines had a differential path length difference and converged to form an end-fire phased array, which emitted light in the plane of the probe. The input wavelength to the device tuned the in-plane emission angle, which steered the beam. The short length of the end-fire waveguide array minimized the inter-waveguide crosstalk at an emitter pitch ($d$) of about $\lambda$.

Three of the four OPAs (I-III) were designed to operate at blue wavelengths with different slab grating curvatures as shown in Fig.\ref{fig:device}d to tune the steering range in the far-field (see OPA Design Section for more details). For OPA I and III, light from the end-fired phased array emitted into a free-propagation region (FPR) and was subsequently radiated out of the plane of the probe with a slab grating at the end of the FPR. The FPR was introduced to transition between the end-fire phased array and the curved grating. Despite the phased array pitch not satisfying the $\lambda/2$ criterion, the high-order lobes of the end-fired phased array were cropped by the side walls of the FPR as illustrated in Fig. \ref{fig:device}d. The FPR also improved the steering range of the device, as the OPA designs with the FPR achieved beam steering through two mechanisms illustrated in Fig. \ref{fig:sim}a: (1) the lateral translation of the beam that first formed in the FPR, and (2) the dependence of the out-of-plane emission angle on the incident angle of the FPR beam on the curved grating. For OPA II, with the aim of a reduced device footprint, the end-fired phased array was coupled directly to a straight one-dimensional slab grating. In this design, high-order lobes with sufficiently large emission angles were minimized because they could not propagate along the full length of the slab grating. The design of OPA IV is similar to that of OPA II but was designed to operate in amber wavelengths to demonstrate the applicability of the OPA architecture in other wavelengths used in optogenetics. The four slab gratings had a constant grating period ($\Lambda$), duty cycle of 50\%, and were fully etched.

\begin{table}
\caption{Specifications of design parameters for different OPA Types. }\label{table:design}
\centering
\begin{tabular}{ccccccccc}
\hline
OPA & $\lambda$ (nm) & SiN Layer & $d$ (nm) & FSR (nm) & \shortstack{\\Slab \\ Curvature} & \shortstack{\\FPR \\ Length (\textmu m)} & \shortstack{$\Lambda$ (nm)} \\
\hline
I & 460-484  & \shortstack{\\120 nm \\ LPCVD} & 400  & 24.5 & Concave & 100 & 480  \\
II & 450-468 & \shortstack{\\150 nm \\PECVD} & 400 & 19 & Straight & No FPR & 400  \\
III & 460-478 & \shortstack{\\150 nm \\PECVD} & 400  & 19 & Convex & 40 & 480   \\
IV & 574-598 & \shortstack{\\200 nm \\LPCVD} & 500  & 25 & Straight & No FPR & 670  \\
\hline
\end{tabular}
\begin{tablenotes}
\item $d$ is the pitch between waveguide emitters in the end-fire array. \item FSR is the calculated free spectral range.
\item FPR is the free propagation region in the slab grating.
\item $\Lambda$ is the period of the slab grating.
\end{tablenotes}
\end{table}


\subsection{OPA Design} \label{sec:OPAdesign}

 \begin{figure}[!ht]
\centering
\includegraphics[width=\linewidth]{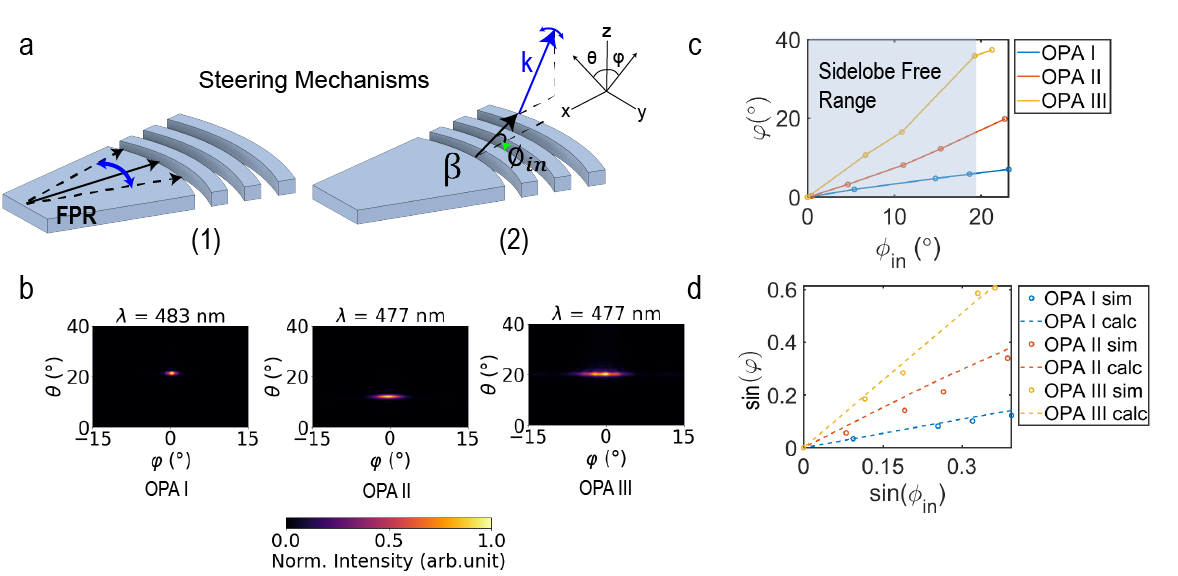}
\caption{\textbf{a} Schematics of the two different steering mechanisms. (1) lateral translation of a beam in the FPR, and (2) the dependence of the out-of-plane emission angle on the incident angle of the emission in the grating. (2) shows the directions of the far-field angles, ($\theta, \phi$) and in-plane angle ${\phi}_{in}$. $\beta$ is the propagation constant of the in-plane emission, $G$ is the grating vector and $k$ is the propagation constant of the output beam. \textbf{b} Simulated 3D FDTD far-field beam profiles for each type of slab grating curvature. The far-field simulations are performed in a medium with refractive index of 1.46, matching that of SiO2. \textbf{c} Comparison of the simulated steering ranges for the different types of slab grating curvatures. The shaded region indicates the side-lobe-free steering range. \textbf{d} The simulated relationship between $\varphi$ and $\phi_{\text{in}}$ is compared with the calculated relationship using Eq. \ref{eqn:steer}.}
\label{fig:sim}
\end{figure}

To intuitively understand the OPA operation, we first make the simplification to assume the optical input and output are plane waves. In this case, the relationship between the in-plane propagation angle, $\phi_{in}$, and out-of-plane emission angles, ($\theta$, $\varphi$), follows the phase-matching condition, 
\begin{subequations}
 \begin{equation}
     \beta \cos \phi_{\text{in}} = k\sin \theta + m_x G_x,
 \end{equation}
 \begin{equation}
     \beta \sin \phi_{\text{in}} = k \sin \varphi + m_y G_y,
 \end{equation}
 \end{subequations}

where $\beta$ is the propagation constant of the in-plane emission, $m_{x(y)}$ is an integer, $G_{x(y)}$ is the $x(y)$ component of the grating vector, and $k$ is the propagation constant of the output beam as depicted in Fig. \ref{fig:sim}a. For the first order beam emission, by setting $m = 1$, we can simplify the $y$ component to be 
\begin{subequations}
    \begin{equation}
\sin \varphi = \Gamma \sin \phi_{\text{in}},
\label{eqn:steer}
\end{equation}
\begin{equation}
\Gamma = 
\begin{cases}
\frac{1}{k} \left(\beta - \frac{2\pi}{\Lambda}\right), & \text{for OPA Type I}\\
\frac{\beta}{k},  & \text{for OPA Type II}\\
\frac{1}{k} \left(\beta + \frac{2\pi}{\Lambda}\right), & \text{for OPA Type III}
\end{cases}
\end{equation}
\end{subequations}
where $\Gamma$ depends on the curvature of the slab grating. 

We verified the OPA designs with three-dimensional (3D) finite difference time domain (FDTD) simulations. Examples of the simulated far-field beam profiles for each OPA are shown in Fig.\ref{fig:sim}b. These simulations confirmed the efficacy of the slab grating designs with the end-fire phased array in eliminating higher-order emissions. As the input wavelength is tuned, emitted beams are steered in $\varphi$, with small changes of $<{2.5}^{\circ}$ in $\theta$. A comparison of the simulation results, highlighting the steering capabilities of the three  OPA designs in $\varphi$ for a given ${\phi}_{in}$, is shown in Fig. \ref{fig:sim}c. There is a trade-off between the beam width and the steering range. While a convex grating curvature has the largest steering range, the width of the beam is also magnified, resulting in approximately the same number of resolvable points in the far-field for the three OPA designs. Figure \ref{fig:sim}d shows that the simulated relationship between ${\phi}_{in}$ and $\varphi$, agrees well with the values calculated using Eq. \ref{eqn:steer}.

\subsection{Beam Profile Characterization}

\begin{figure}[!ht]
\centering
\includegraphics[width=\linewidth]{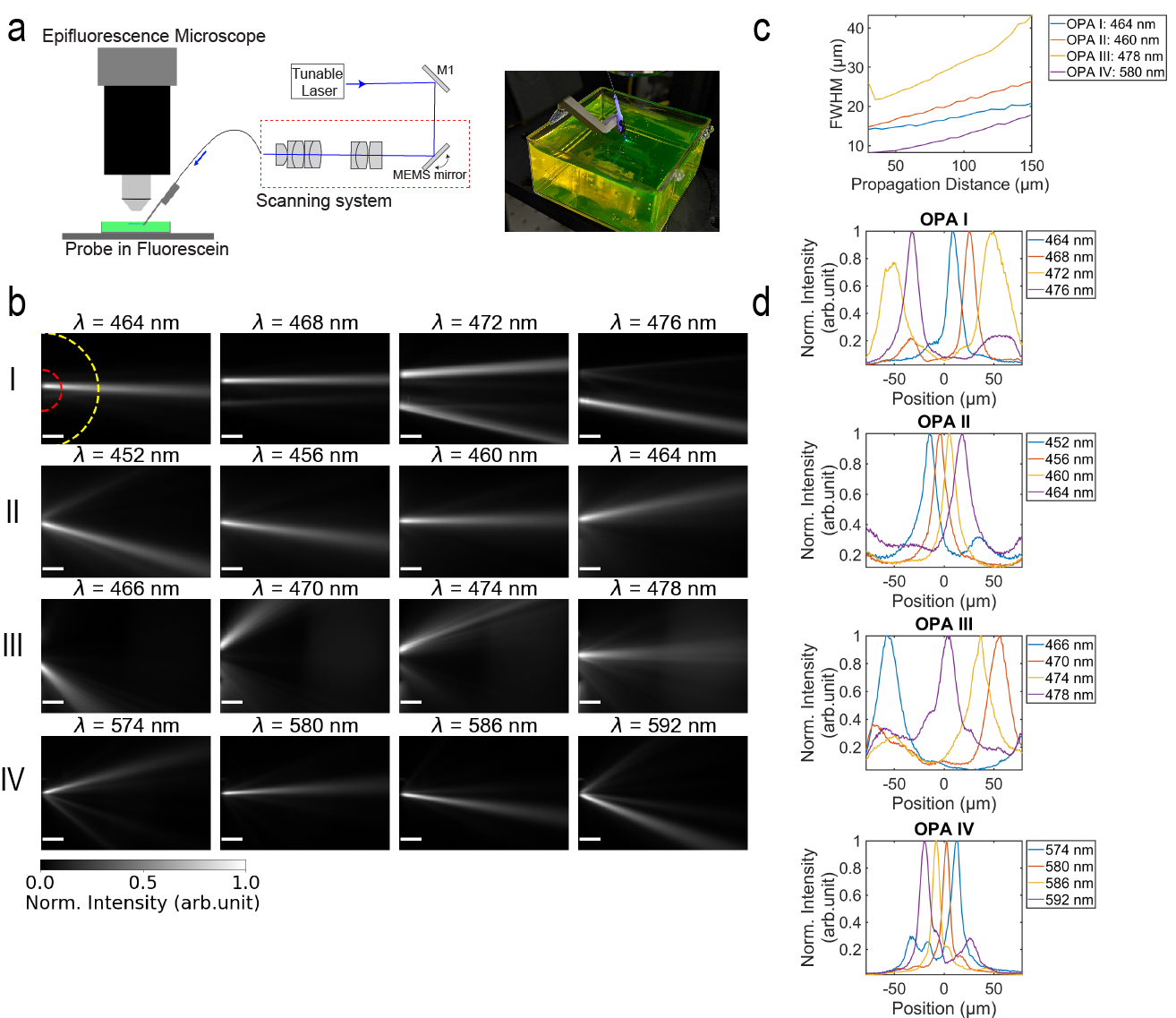}
\caption{\textbf{a} Diagram of the measurement setup showing a neural probe connected to a scanning system and photograph of a probe emitting blue light in a 100 \textmu M fluorescein fluorescent dye solution. \textbf{b} Measured top-down intensity beam profiles of the four  OPA designs in solution at various wavelengths ($\lambda$). The scale bars are 50 \textmu m. For each OPA, we characterized beam-steering along an arc of 50 \textmu m radius (annotated in red) and 140 \textmu m radius (annotated in yellow). \textbf{c} Top-down measured FWHM beam widths versus propagation distance for the four OPA designs at a single wavelength. \textbf{d} Radial line profiles of
the images in \textbf{b} at a propagation distance of 50 \textmu m from the OPA. The line profiles are plotted as the intensity versus lateral position of the steerable beams along the arc with a 50 \textmu m radius.
}
\label{fig:fluor}
\end{figure}

To characterize the optical beam and steering properties of the OPAs, we measured the emission profiles of the fabricated neural probes in non-scattering media. Each probe was inserted into a chamber of fluorescent dye solution corresponding to its operating wavelength (450-484 nm: fluorescin or 574-598 nm: Texas Red). When immersed in solution, the probes were angled such that the emitted beams were approximately parallel to the surface of the liquid. The fluorescence signal was captured with an epifluorescent microscope illustrated in Fig. \ref{fig:fluor}a. For the optical input, a laser scanning system similar to the one described in \cite{Chen2023} was used to address light to individual OPAs via a custom multicore fiber attached to the probe \cite{Azadeh2022}. More details of the measurement setup are described in the Methods section. A photograph of a packaged neural probe with an OPA emitting blue light in fluorescein solution is shown in Fig. \ref{fig:fluor}a. The top-down fluorescence images of the beam profiles for the OPA designs at multiple wavelengths are shown in Fig. \ref{fig:fluor}b. Each OPA was tuned across its full free spectral range (FSR) (close to 20 nm) by using a supercontinuum laser coupled to a tunable bandpass filter. While the figure illustrates only a selection of wavelengths, the emitted beams can be continuously steered throughout the entire FSR, as demonstrated in Supplementary Videos 1 and 2. Approximately $0.2 - 2.5$ \textmu W was emitted from the device under test.

The beam profiles measured in fluorescent dyes, shown in Fig. \ref{fig:fluor}b, capture changes in emission angle along the $\varphi$-axis as well as additional lateral movements in the central position of the beam due to propagation in an FPR. Table \ref{tab:results} summarizes the device footprint and the steering results of the four OPA devices in fluorescent dye. In the context of simultaneous optogenetic stimulation and electrophysiological recording, we 
report the beam steering range along arcs with a radius of 50 \textmu m and 140 \textmu m centered on the emitting region of the OPA. These propagation distances correspond to the estimated electrode detection range for large amplitude spikes (> 60 {\textmu}V) and spike amplitudes above background noise \cite{Henze2000}. The single-lobe steering ranges reported in Table \ref{tab:results} have a peak beam intensity-to-background ratio between 3 and 20 times ($\gtrsim$ 4 dB).  This background suppression level is sufficient to reliably induce spikes in the areas targeted by the main beam on neurons expressing ChR2 with spiking probability > 70\%, while minimizing the light-induced spiking probability in untargeted regions to < 35\% \cite{Jackman2014}. The single-lobe steering range for the four OPA designs covers at least 60\% of the total beam steering range of the OPA (72$\%$, 63$\%$, 87$\%$, and 81$\%$ for OPA I-IV, respectively).

When comparing steering performance among OPAs designed for blue wavelengths, OPA Type I stands out with the lowest beam divergence and the narrowest beam width of approximately 20 \textmu m at a 140 \textmu m propagation distance, as shown in Fig. \ref{fig:fluor}c. In contrast, OPA Type III shows the widest single-lobe steering range enhanced by the negative curvature of the grating slab, as shown in Fig. \ref{fig:fluor}d and reported in Table \ref{tab:results}. Assuming that the length along a 180$^\circ$ arc represents the maximum steering range, the single-lobe steering range of OPA Type III covers 71 $\%$ and 53$\%$ of the maximum range at propagation distances of 50 \textmu m and 140 \textmu m, respectively. Regarding the number of resolvable points, both OPA Types I and III achieved approximately 4-5 resolvable points over the single-lobe steering range, while OPA Type II had the lowest spatial resolution near the probe. For comparison, when the optogenetic stimulation is limited by the cell size \cite{Lees2024}, the maximum number of achievable resolvable points along the 180$^\circ$ arc length is about 10 and 29 (assuming a cell diameter of 15 {\textmu}m), respectively, for propagation distances of 50 \textmu m and 140 \textmu m. This translates to OPA III capturing approximately 46 $\%$ of the maximum resolvable points at 50 \textmu m, but only 18 $\%$ at 140 \textmu m. Although OPA Type II has the worst steering performance near the probe due to the lack of FPR, the design is advantageous for neural probes with dense OPA emitters, as it has the most compact footprint. Furthermore, OPA Type II can achieve larger steering ranges than OPA Type I at longer propagation distances $\ge$ 150 \textmu m, resulting from the predominant influence of the grating curvature on beam steering (see \ref{supp:200}).

\begin{table}[h!]
    \begin{center}
    \caption{Comparison of the OPA designs in fluorescein solution.}    \label{tab:results}
    \resizebox{\textwidth}{!}{%
    \begin{tabular}{|c|c|c|c|c|c|c|}
        \hline
        OPA & $\lambda$ (nm) & \shortstack{\\ Device \\ Size (mm$^{2}$)} & \multicolumn{2}{c|}{\(L_{prop}\) = 50 \textmu m} & \multicolumn{2}{c|}{\(L_{prop}\) = 140 \textmu m}\\
        \cline{4-7}
        & & & \shortstack{\\Single-lobe \\ Steering Arc Length (\textmu m)} & \shortstack{Beam width\\FWHM (\textmu m)} & \shortstack{\\Single-lobe \\ Steering Arc Length (\textmu m)} & \shortstack{Beam width\\FWHM (\textmu m)} \\
        \hline
        I & 460-484 & 0.018 & 66.4 & 14.4 $\pm$ 2.5* & 79.8 & 17.5 $\pm$ 2.3* \\
        II & 450-468 & 0.007* & 32.9  & 19.7 $\pm$ 3.1  & 78.2 & 29.7 $\pm$ 4.3 \\
        III & 460-478 & 0.011 & 113.0*  & 24.5 $\pm$ 6.6 & 235.4* & 44.0 $\pm$ 14.3 \\
        \hline\hline
        IV & 574-598 & 0.006 & 33.0 & 9.9 $\pm$ 1.4 & 86.7  & 18.1 $\pm$ 2.4 \\
        \hline 
    \end{tabular}%
    }\\
    \end{center}
    \begin{tablenotes}
        \item{*}The best values for blue wavelengths. 
\end{tablenotes}
    \end{table}

\begin{figure}[!ht]
\centering
\includegraphics[width=\linewidth]{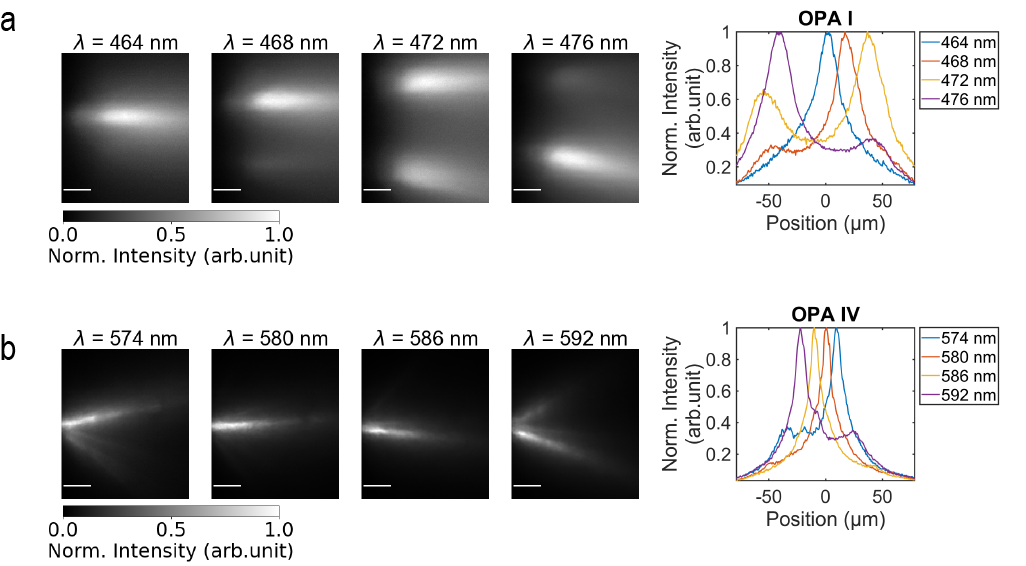}
\caption{ Measured intensity top-down beam profiles of \textbf{a} OPA I and \textbf{b} OPA IV in fixed tissue. The scale bars are 25 \textmu m. On the right are radial line profiles of the measured images at a propagation distance of 50 \textmu m showing the intensity versus lateral position of emitted beams as they are steered in tissue.}
\label{fig:tissue}
\end{figure}

We also characterized beam-steering in fixed tissue. We implanted two of the probes, one with Type I OPA for blue light and the other with Type IV OPA for amber light, into fixed brain slices from wild-type mice that were 2 mm thick and stained with the corresponding fluorescent dye. The probes were inserted at shallow depths such that the emitting OPAs were implanted $< 100$ \textmu m from the tissue surface. Figure \ref{fig:tissue} shows the beams formed in the tissue. At blue wavelengths, the beam width (FWHM) after propagating 50 \textmu m was broadened by 18.7 \textmu m to an average of 33.1 \textmu m. At amber wavelengths, the mean FWHM of the emitted beams from OPA IV was 18.6 \textmu m, which was 8.7 \textmu m more than in non-scattering media. The broadening of the beam within the tissue effectively reduced the number of achievable resolvable points from 4-5 to approximately 2.

\subsection{\emph{In Vivo} Experiments}

\begin{figure}[!ht]
\centering
\includegraphics[width=\linewidth]{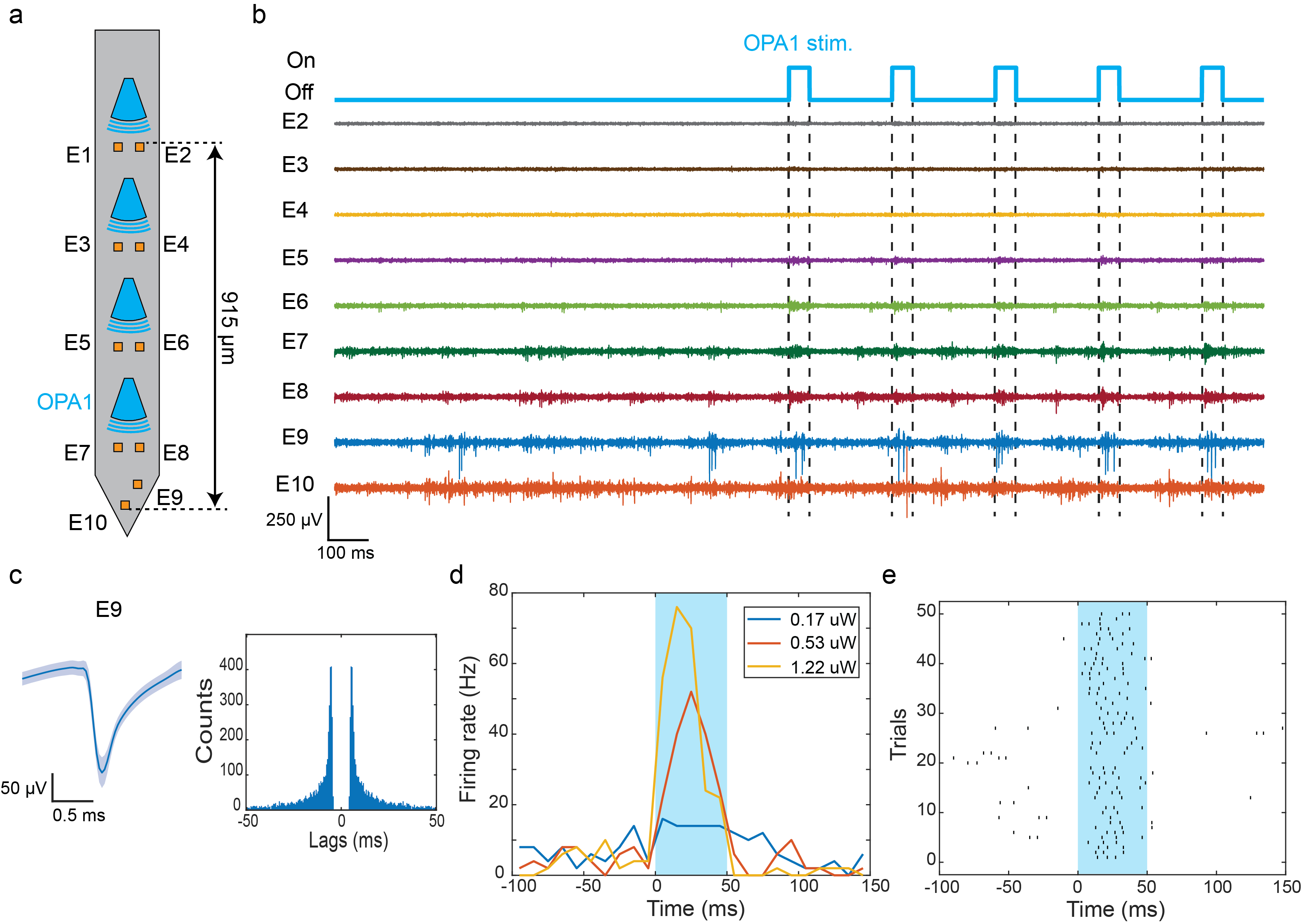}
\caption{Demonstration of optogenetic stimulation with the OPA Type I emitter in \emph{in vivo} awake head-fixed mice.\textbf{a} Layout of the OPA probe used in the \emph{in vivo} experiment. It contained 4 OPA Type I emitters and 10 microelectrodes. \textbf{b} Snapshot of the filtered electrophysiological recording (common average referencing (CAR) + bandpass filtered (300-6000 Hz) + artifact subtraction). OPA1 labeled in \textbf{a} was used for optogenetic stimulation. The trigger signals of the optical pulses are indicated above, with dashed lines marking the onset and offset of the pulse. Electrode indices are labeled on the side. The output power was 1.22 \textmu W. \textbf{c} Waveform and autocorrelogram of a sorted unit recorded on electrode E9. The solid blue line represents the mean waveform and the shaded grey area represents the standard deviation of the waveform. \textbf{d} The mean peristimulus time histogram (n = 50 trials) of the unit in \textbf{c} under varying optical stimulation power from OPA1. \textbf{e} Raster plot of the unit recorded with electrode E9 during optical stimulation from OPA1 in 50 trials, each consisting of 50-ms optical pulse stimulation at a power level of 1.22 \textmu W.}
\label{fig:fig5_invivo}
\end{figure}

To validate the photostimulation \emph{in vivo}, we implanted a hybrid probe with 4 OPA Type I emitters and 10 microelectrodes, as shown in Fig. \ref{fig:fig5_invivo}a, in the cortex of Thy1-ChR2-YFP mice. We targeted probe insertion into Layer V of the motor cortex as the brain region has a high level of expression of the ChR2 channelrhodopsin \cite{Wang2007}. A snapshot of the recording trace before and during a photostimulation pulse train across 9 electrodes is presented in Fig. \ref{fig:fig5_invivo}b (one electrode was not electrically connected). A 4 Hz pulse train with a 50-ms pulse width was applied for 2.5 s, repeated five times with a 10 s interval for recovery, resulting in a total of fifty 50-ms stimulation pulses. Subsequent spike sorting of the recording verified that the stimulated unit exhibited a clean spike waveform and minimal refractory period violation in its autocorrelogram (see Fig. \ref{fig:fig5_invivo}c). 

Increasing the optical output power from the OPA emitter resulted in a higher average stimulated firing rate and a reduced spike latency of the first spike after the onset of the optical pulse in the stimulated unit, as presented in Fig. \ref{fig:fig5_invivo}d. Across the power output range of 0.17 to 1.22 \textmu W (0.35 to 2.54 mW/mm$^{2}$ estimated with the 1/e$^{2}$ beam size measured in fluorescein data), the average firing rate increased from 14 to 50 spikes/s, while the average spike latency shortened from 24 ms to 14 ms. The raster plot in Fig. \ref{fig:fig5_invivo}e demonstrates a consistent stimulation response over all 50 pulse repetitions, with a 94\% probability of detecting at least one spike within the optical pulse.  

We have attempted to detect a difference in photostimulated spiking patterns as we steered the beam. However, the small steering angle constrained by the laser wavelength tuning range available at the \emph{in vivo} experimental site yielded no discernible differences in the firing patterns across the electrodes. The laser wavelength tuning range of 485.5 to 489 nm resulted in a beam translation of $\sim $ 9 \textmu m at 50 \textmu m  propagation distance, less than the size of a soma body in the motor cortex (10 - 20 \textmu m \cite{Oswald2013}). The broadening of the beam in the tissue observed in Fig. \ref{fig:tissue}a further reduced the spatial resolution of the beam. Also, since the star couplers were not coated with optically opaque epoxy to prevent an increase in the size of the device in the animal experiments, the light that scattered from the star coupler (positioned at $\approx$ 140 \textmu m from electrode E9) could have contributed to the observed photostimulated response in the experiment and broadened the stimulation volume. Despite these limitations, the experiment demonstrated that a neural probe with an array of OPAs and electrodes can perform simultaneous photostimulation and electrophysiological recording \emph{in vivo}. 



\section{Discussion and conclusion}

In this study, we conducted a comprehensive characterization of four sidelobe-free OPA designs for optogenetic stimulation applications, with three designs using different grating curvatures (concave, straight, and convex) for blue wavelengths and one straight slab grating for amber wavelengths. The FPR for OPA Type I and III introduces an additional translation of the lateral beam at the grating emission site, increasing the steering range and the resolvable points close to the probe within the electrode detection range  (< 140 \textmu m). In contrast, OPA Type II with a straight slab grating and no FPR slab compromises the steering performance near the probe for a more compact footprint, making it suitable for probe designs requiring high-density integration of OPA emitters. \emph{In vivo} experiments demonstrated that the OPAs emitted sufficient power for effective optogenetic stimulation in Thy1-ChR2 mice, and the neuronal spiking activity was recorded by the electrodes on the probe.

The proof-of-concept experiments show the need for continued design improvements to achieve spatially selective optogenetic stimulation in the deep brain with steerable beams in the future. The beam broadening in tissue and decreased peak-to-background ratios reduced the number of resolvable spots. However, we found that the beam profiles measured in the mouse cortex were significantly larger than the beam width predicted in the scattering simulation presented in Section \ref{supp:scattering_sim}. We speculate that one reason for this difference is the use of fixed tissue, which can have a higher optical scattering than fresh samples, as reported in \cite{Silvestri2016, Markus2023}. Further evidence of this hypothesis is provided by the improved agreement between the experimental and simulated beam widths when using higher scattering coefficient settings in the simulation. In addition, we observed that, in simulation, the beam profile captured by the microscope can overestimate the actual beam width in tissue due to the scattered fluorescence signal caused by the tissue layer covered on top of the OPA, similar to the results shown in \cite{xue2024implantable}. To validate these potential effects observed in the simulation, future work could collect the cross-sectional beam profile propagating through fresh brain slices with predefined thicknesses as described in \cite{Xue2022}. To reduce the beam broadening and improve the stimulation resolution, future experiments should use red-shifted opsins and red wavelengths, which have lower optical scattering in tissue compared to blue wavelengths. Furthermore, generating a steerable focused beam \cite{Shirao2022} would reduce the broadening of the beam and localize the stimulation at the focal point where the intensity is the highest.

To conclusively show spatially selective optogenetic stimulation \emph{in vivo}, several strategies can be pursued. In addition to managing the beam width by designing focusing gratings, 

using a more widely wavelength tunable laser would increase the steering range. Blocking stray and scattered light with light-absorbing epoxy would improve the confinement of the effective photostimulation volume. On the chip, scattered light from the photonic circuit components, such as the star coupler, can be blocked using the top metal routing layers. On the biological side, using soma-targeted opsins would spatially restrict the opsin expression in the neurons to reduce the probability of stimulating the neurons via their axons or dendrites \cite{Shemesh2017, Chen2021}. Lastly, instead of monitoring the effects of optogenetic stimulation with electrophysiological recordings, switching to all-optical brain interrogation, where neural activity is recorded by calcium or voltage imaging \cite{Packer2015, Hochbaum2014}, may enable more direct and higher resolution observations of the spatial differences in neural activity elicited from a scanning optical beam.

In summary, we have reported passive single-lobe steering OPAs for blue and amber wavelengths on Si neural probes for \emph{in vivo} optogenetic applications. A unique photonic circuit architecture that combines an in-plane phased array with a light-emitting grating enabled single-lobe beam steering in an ultracompact footprint without meeting the half-wavelength pitch criterion. Beam steering photonic circuits on neural implants bring dynamically defined light patterns into deep brain regions to enable a new class of neural circuit activity mapping experiments.

\section{Methods}

\subsection{Device fabrication}

The neural probes with OPAs were fabricated by Advanced Micro Foundry (AMF) on 200-mm diameter Si wafers. Initially, SiO$_{2}$ and SiN were deposited on the Si wafer for the waveguide cladding and core. The choice between LPCVD or PECVD for SiN depended on the required refractive index for specific OPA designs. The thickness and type of SiN for each OPA design is described in Table \ref{table:design}. Subsequently, 193-nm Deep Ultraviolet (DUV) photolithography and reactive ion etching were applied to the SiN layer to pattern the photonic circuits. Then, SiO$_{2}$ was deposited above the SiN to encapsulate the waveguide core within the SiO$_{2}$ cladding. Three layers of aluminum (Al) routing and TiN surface electrodes were deposited above the waveguide structure for electrophysiological recording electrodes and metal wiring. The outline of the probe was defined using deep trench etching from the front of the wafer, and the probes were released from the wafer by backgrinding the wafer to a thickness of about 100 \textmu m. An additional laser roughening treatment process, detailed in \cite{Chen2023}, was performed on the electrodes of each probe to reduce the electrode impedance to below 2 M$\Omega$, an impedance threshold to achieve an acceptable signal-to-noise ratio (SNR) for electrophysiological recording \cite{Neto2018}.

\subsection{Neural probe packaging}

The neural probes were packaged on different substrates depending on their intended use. The packaging procedure was similar to the methods described in \cite{Chen2023}. For beam profile measurements, the probes were bonded to metal holders using heat-curable metal epoxy (Ablebond 84-1LMIT1, Loctite, Stamford, CT, USA). Prior to attaching the multi-core fiber (MCF) to the chip, a small amount of optically opaque epoxy (EPO-TEK-320, Epoxy Technology, Billerica, MA, USA) was applied to the star coupler on the OPA emitters to block stray light from the device. A custom 10- or 16-core MCF fiber was then actively aligned to the probe edge couplers using a 5-axis fiber alignment stage with a fiber rotator for rotational alignment. After achieving optimal alignment between the MCF and the chip, we incrementally applied small amounts of low-shrinkage UV-curable epoxies (OP-67-LS and OP 4-20632, Dymax Co., Torrington, CT, USA) to the probe and MFC, followed by UV curing with a UV LED system (CS2010, Tholabs, Newton, NJ, USA) after each drop to minimize alignment drift due to epoxy shrinkage. After the MCF was secured with epoxy, an additional 5 minutes of epoxy was applied to the back of the probe holder and MCF to provide stress relief. Finally, optically opaque epoxy was applied to cover the probe base and fiber to prevent stray light from reaching the samples.

For the probe used in the \emph{in vivo} animal experiment, we mounted the probe on a custom printed circuit board (PCB) with a metal block added to the bottom of the probe to address clearance issues between the MCF and the base of the PCB. The probe was bonded to the metal block and PCB with the heat-curable metal epoxy. The probe was then wire bonded to the PCB with Au wires. To minimize light-induced artifacts in electrophysiological recordings, a UV-curable encapsulating epoxy (Katiobond GE680, Delo, Germany) and optically opaque epoxy were applied to minimize stray light from the fiber reaching the wire bonds. Similar MCF-to-chip attachment procedures described above were then performed after electrical packaging, as optical packaging requires higher alignment accuracy. We did not use the optically opaque epoxy to cover the star coupler for the probe used for \emph{in vivo} experiments, as it increases the size of the implant. Finally, the probes were left undisturbed for at least 12 hours to allow the epoxy to fully cure before removal from the packaging assembly.

\subsection{Optical scanning system and tunable laser sources}

We used a custom-built optical scanning system, described in detail in \cite{Chen2023} to address the emitters on the probe with light from an external laser source. This system used a micro-electromechanical system (MEMS) mirror and custom optical lenses to direct and focus the input light onto each fiber core of the MCF (core size of $\sim$ 2.6 \textmu m). The insertion loss between the input of the scanning system and the MCF output was approximately 2-3 dB. We used a program scripted in MATLAB to search for the MEMS coordinates corresponding to the fiber cores on the MCF and control the MEMS during the experiments. The scanning system used a fiber input port for adaptability to various laser sources. For beam profile characterization in fluorescent solutions and mouse brain slices, a supercontinuum white laser (NKT SuperK Fianium) with a tunable filter covering a range of 400 to 1100 nm was employed as the input light source. For \emph{in vivo} experiments, an external cavity laser with motorized wavelength control was used to tune the wavelength between 484.3 and 491 nm (TOPTICA Photonics Inc., DLC DL pro tunable laser system with integrated fiber coupler). We constructed a free-space-to-fiber coupling stage for each laser source to couple light into a single-mode fiber (460-HP, Nufern Inc.), which was then connected to the scanning system. An optical shutter and a variable neutral density filter were added in between the laser and the fiber-coupling stage for gating the input beam and controlling the optical power.

\subsection{Animals}

The measurements in fixed brain slices were carried out at the Max Planck Institute of Microstructure Physics in Halle, Germany. The \emph{in vivo} experiments were conducted at the Krembil Brain Institute in Toronto, Canada. Fixed brain slices were provided by the Fraunhofer Institute for Cell Therapy and Immunology and the surgical procedures to extract the brain were performed in accordance with German laws. The \emph{in vivo} experimental procedures described here were reviewed and approved by the animal care committees of the University Health Network in accordance with the Canadian Council on Animal Care guidelines. Adult male and female Thy1-ChR2-YFP mice (The Jackson Laboratory, Bar Harbor, Maine, stock number 007612) were kept in a vivarium maintained at 22 °C with a 12-h light on/off cycle. Food and water were available ad libitum.

\subsection{Preparation of fixed mouse brain slices stained with fluorescent dye}

To assess the beam profile emitted from each OPA type within a scattering medium, we captured beam propagation in fixed brain tissue slices. We used the whole brain extracted from wild-type mice (C57BL) at the age of 30--150 postnatal days. The whole brain samples were immersed in 1.5--2 \% paraformaldehyde (PFA) for 8 to 12 hours for fixation. Then they were stored in 1x phosphate-buffered saline (PBS) at 4$^{\circ}$C before use. Before experiments, whole brains were sectioned into 2-mm-thick coronal slices using a brain matrix (Alto Brain Matrix stainless steel 1 mm mouse coronal 45-–75 gm; Harvard Apparatus) and stirrup-shaped blades (Type 102, Carl Roth GmbH + Co. KG, Karlsruhe, Germany). The slices were then permeabilized with 0.3\% Triton X solution for 30 minutes, followed by three 5-minute washes in PBS. The slices were then immersed overnight in 100 \textmu M fluorescein or Texas Red fluorescent dye, depending on the operating wavelength of the OPA being tested (450--484 nm: fluorescein,  574--598 nm: Texas Red). The slices were ready for use the next day after three additional 5-minute washes in PBS.

\subsection{Beam profile characterization}

We evaluated the steering range and beam profile of each OPA type in non-scattering and scattering media. The input light was set to TM-polarized light for maximum power output from the OPA grating using an in-line fiber polarization controller. For non-scattering medium, the shank of the probe was immersed in either 100 \textmu M  fluorescein or Texas Red fluorescent dye, depending on the operating wavelength for the OPA type (450--484 nm: fluorescein,  574--598 nm: Texas Red). A 4-axis micromanipulator (uMp-4, Sensapex, Oulu, Finland) precisely controlled the probe's movement while top-down beam profiles were captured using a wide-field epifluorescence microscope (Cerna, Thorlabs) equipped with a 10x objective (M Plan Apo 10$\times$, NA = 0.28, Mitutoyo Deutschland GmbH), appropriate filter cubes (49002, Chroma Technology Corporation, Bellows Falls, VT, USA for GFP and mCherry-C-000-ZERO, Semrock, Rochester, NY, USA for mCherry), and an sCMOS camera (Prime BSI, Teledyne, Photometrics, Tucson, AZ, USA). Additionally, side-view beam profiles were acquired with a microscope to examine the beam profile thickness and potential stray light emission from the probe other than the OPA grating. The side-view microscope consisted of a variable magnification microscope (12X Zoom Lens System, Navitar, Rochester, NY, USA) equipped with a 5x objective, a GFP emission filter (MF525-39, Thorlabs) or a Texas Red emission filter (MF630-69, Thorlabs), and a CMOS camera (Grasshopper3, USB 3.0, Teledyne FLIR, Wilsonville, OR, USA). A supercontinuum white laser (NKT SuperK Fianium) with a tunable filter with a tuning range of 400 to 1100 nm provided the input light source to the probe.

To evaluate beam profiles in scattering media, the probe shank was inserted into fixed brain tissue prepared following the procedures outlined in the Methods section. The same fluorescence microscope system described above was utilized to capture top-down beam profiles within the scattering medium. For capturing the beam profile close to the tissue surface, the OPA emitter was first inserted fully into the tissue, and a series of beam profiles over the complete FSR of the OPA was acquired. This image acquisition process was repeated after retracting the probe at 30-50 \textmu m increments until the OPA emitter was out of the tissue.

\subsection{Experimental procedures for awake head-fixed animal experiments}

In preparation for the awake head-fixed experiment, the headplate was mounted onto the mouse skull 2--4 days prior to the experiment. Thy1-ChR2-YFP mice at the age of 60-90 postnatal days were used. The mouse was first anesthetized with 5\% isoflurane/oxygen by induction and maintained with 1--2\% isoflurane/oxygen. It was then secured in a stereotaxic frame using ear bars (Model 902, David Kopf Instrument, Tujunga, California). Before mounting the headplate, two bone screws (Item No. 19010-10, Fine Science Tools) were implanted for ground and reference electrodes. We followed the same ground and reference configuration described in \cite{Golabchi2021}. The insertion positions (stereotaxic coordinates of AP: 0 to -0.5 mm, ML: 1.2 mm for targeting motor and somatosensory cortex, with the origin aligned to bregma) were also marked using a dental drill. Finally, the headplate was attached to the mouse skull using dental cement (C\&B Metabond, Parkell, Edgewood, NY, USA). The mouse was placed in a separate cage for recovery.

On the day of the experiment, the mouse with a headplate was anesthetized following the same protocol as before. We then used a dental drill to open a circular craniotomy of 1--2 mm diameter at the previously marked position for probe insertions. The dura was left intact to minimize the motion artifact caused by brain pulsation. We then placed the mouse body in a 3D-printed cone under the microscope to restrain mouse movement, and the headplate was attached to the head bar posts. The probe was mounted on a micromanipulator for precise translation and insertion speed control. The reference and ground screws were connected to the corresponding pins on the Intan headstage, and we attached an additional ground wire from the ground plane of the optical table to strengthen the ground connection.

To ensure successful dura penetration, the probe was inserted into the brain at a speed of 0.5--1 mm/s for the first 500--900 \textmu m. Following confirmation of the probe insertion, the probe was retracted by 200--500 \textmu m to allow the brain to relax and was held in this position for 10--15 minutes. Subsequently, the probe was advanced at a slower speed of 1--2 \textmu m/s to minimize tissue damage until a depth of 1--1.4 mm was reached based on the micromanipulator measured insertion depth, where the electrodes near the tip of the shank were positioned in cortical layers V and VI. The probe was rested for approximately 30 minutes before the start of the photostimulation.

Prior to photostimulation, the input wavelength was set to 485.5 or 490 nm, and an electrically driven polarization controller at the laser output was used to set the input light to TM polarization based on the scattered intensity from the probe captured by the microscope camera. For optogenetic stimulation, a 4 Hz pulse train with 50 ms optical pulses was applied for 2.5 s, repeated five times with a 10 s recovery period between each pulse train. 

After the experiment, the probe shank was immersed in 1\% Tergazyme solution (Sigma-Aldrich, St. Louis, MO, USA) for 2 hours, followed by a 10--20 minute wash in Milli-Q water. The probe was air-dried for subsequent experiments.

\subsection*{Electrophysiological data analysis}

The raw electrophysiological data recorded from the Open Ephys acquisition board \cite{Siegle2017} were preprocessed in Python before spike detection. First, we applied common average referencing using electrode channels that detected fewer spikes as the reference signal. The signal was then filtered with a third-order Butterworth bandpass filter (300--6000Hz). To reduce the amplitude of light-induced artifacts, we calculated the average light-induced artifact waveform for each OPA at the same stimulation power and subtracted it from the onset and offset of the light pulse. Additionally, a blanking window of 1 ms preceding and 2.5 ms following the start and end of the optical pulse was applied to minimize the artifact's impact on spike detection accuracy.

For spike detection and clustering, we utilized the Spyking Circus package\cite{Yger2018}, followed by manual curation of spike clusters via the phy GUI interface \cite{Rossant2020}. To address double counting of the same spikes during the manual merging process, we removed one of the paired spikes that have an interspike interval of less than 0.5 ms. The final selected spike clusters follow four criteria: 1) isolation distance $>$ 10, 2) likelihood ratio $\leq$ 0.3 \cite{Schmitzer2005,Libbrecht2018}, 3) SNR  $\geq$ 3 \cite{Suner2005}, and 4) the percentage of spikes with refractory period violation ($<$ 2 ms) $<$ 2 \% \cite{Fiath2019}.

\section*{Data availability}

Raw electrophysiological data, beam profile images, and the source data for the figures are available at: https://doi.org/10.17617/3.1X2HCV .

\section*{Code availability}

All codes used in this research are available from the corresponding authors upon reasonable request.

\putbib

\section*{Acknowledgements}

The authors thank Holger Cynis, Ines Koska, and Stefanie Geißler from the Fraunhofer Institute for Cell Therapy and Immunology for providing mouse brain tissue samples and Andrei Stalmashonak and Hannes Wahn for their assistance in setting up the optical systems.  Additionally, the authors would like to thank Hanne Wahn for the discussions on the scattering simulations. The authors also gratefully acknowledge funding support from the Max Planck Society, Natural Sciences and Engineering Research Council of Canada, and the Canadian Institute of Health Research.

\section*{Contributions}
A.S., Y.J. and J.C.C.M. conceived the design approach. F.D.C., A.S., and Y.J. performed the device simulations, designed the probes with inputs from W.D.S. and J.K.S.P., and A.S., T.X., and A.G. laid out the designs. H.C., X.L., and P.G.Q.L. were responsible for the wafer fabrication. Devices were packaged by F.D.C., A.S., and T.X.. A.S., F.D.C. and W.D.S. characterized the devices. M.G.K.B prepared the fixed tissue samples. F.D.C., A.S., H.M.C., M.M. conducted the animal experiments. F.D.C., A.S., and J.K.S.P. analyzed the data. A.S., F.D.C., and J.K.S.P. co-wrote the manuscript with inputs from other co-authors. The project was completed under the supervision of T.A.V., W.D.S., and J.K.S.P..

\section*{Supplementary information}

\textbf{Supplementary Video 1} \\
OPA Type I beam-steering in fluorescein solution (452--464nm). The intensity of each frame is adjusted by a constant factor to improve visibility. The video is in real-time. The video is available at: https://doi.org/10.17617/3.1X2HCV \\

\textbf{Supplementary Video 2} \\
OPA Type IV beam-steering in brain tissue slice stained with Texas Red fluorescent dye (574--596 nm). The intensity of each frame is adjusted by a constant factor to improve visibility. The video is in real-time. The video is available at: https://doi.org/10.17617/3.1X2HCV \\

\end{bibunit}
\clearpage

\pagebreak

\begin{bibunit}

\begin{center}
\textbf{\large Single-lobe beam steering in tissue with optical phased arrays on implantable neural probes: supplementary document}
\end{center}



\begin{center}

  Fu Der Chen\textsuperscript{1,2,3,*},
  Ankita Sharma\textsuperscript{1,2,3,*,$\dagger$},
  Tianyuan Xue\textsuperscript{1,2},
  Youngho Jung\textsuperscript{1},
  Alperen Govdeli\textsuperscript{1,2},
  Jason C. C. Mak\textsuperscript{1},
  Mandana Movahed\textsuperscript{4},
  Homeira Moradi Chameh\textsuperscript{4},
  Michael G. K. Brunk\textsuperscript{1},
  Xianshu Luo\textsuperscript{5},
  Hongyao Chua\textsuperscript{5},
  Patrick Guo-Qiang Lo\textsuperscript{5},
  Taufik A Valiante\textsuperscript{2,3,4,6,7},
  Wesley D. Sacher\textsuperscript{1,3},
  Joyce K. S. Poon\textsuperscript{1,2,3,$\dagger$}
\end{center}

\begin{center}
  \textsuperscript{1}\textit{Max Planck Institute of Microstructure Physics, Weinberg 2, 06120 Halle, Germany} \\
  \textsuperscript{2}\textit{Department of Electrical and Computer Engineering, University of Toronto, 10 King's College Road, Toronto, Ontario M5S 3G4, Canada} \\
\textsuperscript{3}\textit{ Max Planck-University of Toronto Centre for Neural Science and Technology}\\
  
  \textsuperscript{4}\textit{ Krembil Brain Institute, University Health Network, Toronto, Ontario, Canada} \\
  \textsuperscript{5}\textit{Advanced Micro Foundry Pte Ltd, 11 Science Park Road, Singapore Science Park II, 117685, Singapore} \\
  \textsuperscript{6}\textit{Division of Neurosurgery, Department of Surgery, Toronto Western Hospital, University of Toronto, Toronto, Ontario, Canada} \\
  \textsuperscript{7}\textit{Institute of Biomedical Engineering, University of Toronto, Toronto, Ontario, Canada}

\textsuperscript{*}These authors contributed equally to this work.

  \textsuperscript{$\dagger$} ank.sharma@mail.utoronto.ca, 
  \textsuperscript{$\ddagger$}  joyce.poon@mpi-halle.mpg.de

\end{center}

\setcounter{equation}{0}
\setcounter{figure}{0}
\setcounter{table}{0}
\setcounter{page}{1}
\setcounter{section}{0}

\makeatletter
\renewcommand{\theequation}{S\arabic{equation}}
\renewcommand{\thefigure}{S\arabic{figure}}
\renewcommand{\thesection}{S\arabic{section}}
\renewcommand{\bibnumfmt}[1]{[S#1]}
\renewcommand{\citenumfont}[1]{S#1}

\section{Steering Arc Length vs. Propagation Distance}
\label{supp:200}

\begin{figure}[!ht]
\centering
\includegraphics[width=0.6\linewidth]{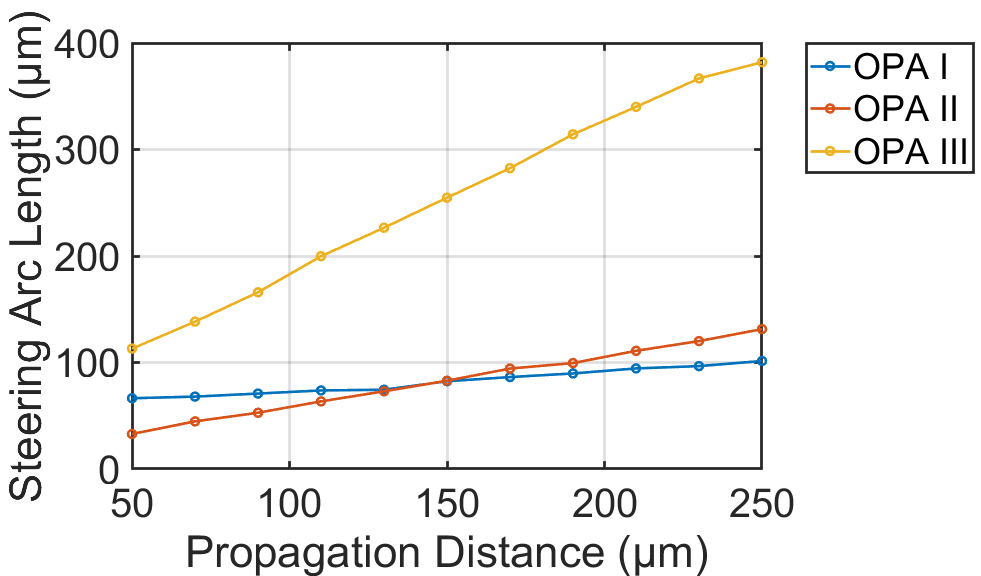}
\caption{Plot of steering arc length as a function of propagation distance (distance away from the OPA). We compare the steering arc length across the three blue OPA designs in the main manuscript (I, II, III) each with a different grating curvature.}
\label{fig:supp_200}
\end{figure}

In Fig. \ref{fig:supp_200}, we plot the sidelobe free steering arc length as a function of propagation distance. For distances $\le$ 150 \textmu m, the steering arc length of OPA I exceeds OPA II. However, OPA II achieves a greater steering range at larger propagation distances as expected from the relationships derived by the phase matching condition for a straight instead of concave grating. Across all propagation distances OPA III achieves the greatest sidelobe steering arc length.

\section{Scattering simulation}
\label{supp:scattering_sim}

\begin{figure}[!ht]
\centering
\includegraphics[width=\linewidth]{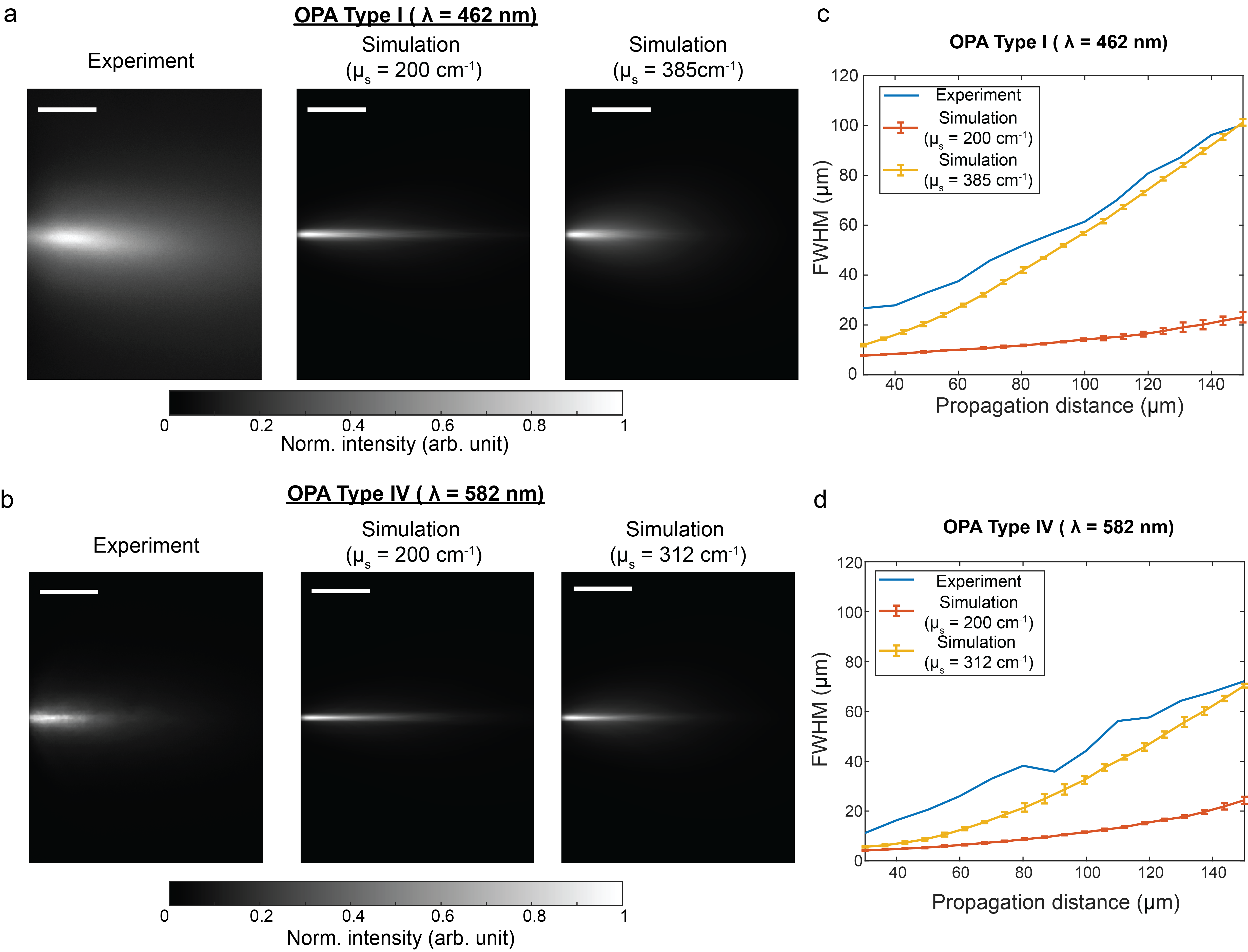}
\caption{Comparison of the simulated and experimental beam profiles for OPA Type I and IV in tissues. Experimental beam profile in fluorescent dye stained fixed mouse cortex slices and the simulated beam profiles at two different scattering coefficients for OPA \textbf{a} Type I and \textbf{b} Type IV. The scale bars are 50 \textmu m. The FWHM beam width measured along the beam propagation distance of the beam profiles from \textbf{a} and \textbf{b} for OPA \textbf{c} Type I and \textbf{d} Type IV.  The scattering simulations were repeated five times for each scattering setting, with each trial generated with new phase masks. The average beam widths (FWHM) were plotted, with error bars representing their standard deviations.}
\label{fig:scattering_simvsexp}
\end{figure}

To simulate the beam profile in tissue, we employed the beam propagation method (BPM) described in \cite{Li2019, Sacher2022}. This method models beam propagation using the angular spectrum method. A phase mask with small phase variations is introduced after each propagation step in the BPM to emulate the index variation across tissues, leading to beam scattering. The optical scattering properties of the medium can be adjusted by designing the phase masks following the design strategy detailed in \cite{Li2019}. The input fields of the OPAs to the BPM were extracted from the 3D FDTD simulation performed in Lumerical.

After obtaining the 3D beam intensity through the scattering simulation, we applied blurring filters to the transverse planes of the 3D beam, the planes parallel to the beam propagation direction, to emulate two beam broadening effects captured by the optical microscope: 1) the out-of-focus light away from the imaging focal plane and 2) the scattering of the fluorescent signal propagating through the top tissue layer covered on top of the OPA. We simulated this effect by convolving the transverse planes of the 3D beam intensity with a set of degraded point spread functions (PSF) obtained using a similar method described in \cite{xue2024implantable}. The degraded PSF for each traverse plane was generated by forward propagating a diffraction-limited Gaussian point source \cite{saleh2007fundamentals}(calculated with the 10$\times$ objective used in the experiment) with scattering to the tissue surface and then propagating in reverse direction to the predefined in-focus object plane. This reversed beam propagation approximates the point spread function detected by a microscope with a magnification of one.  Any transverse planes located above the defined tissue surface were convolved with the PSF generated at the tissue surface assuming that light was incident on the surface. Subsequently, the processed transverse planes were summed together to form the simulated 2D images shown in Figs. \ref{fig:scattering_simvsexp}a and b. To match the scattering properties of the mouse cortex, we set the scattering coefficient (\textmu$_{s}$), absorption coefficient (\textmu$_{a}$), and anisotropy factor (g) to 200 cm$^{-1}$, 0.62 cm$^{-1}$, and 0.86, respectively \cite{Yona2016}. We also assumed that the in-focus object plane of the beam is positioned at 60 \textmu m in depth within the tissue. 

Figures \ref{fig:scattering_simvsexp}a and b compare the experimental and simulated beam profiles for OPA Type I and IV. However, significant discrepancies in FWHM beam width were observed between the simulated and the experimental results for the case with scattering properties that correspond to the mouse cortex (Figs. \ref{fig:scattering_simvsexp}c and d). Better agreement was achieved by increasing the scattering coefficient to 385 cm$^{-1}$  and 312 cm$^{-1}$ for OPA Types I and IV. This result suggests that the optical scattering in the prepared fixed brain slices may be higher than the nominally reported value for fresh brain slices \cite{Juboori2013,Yona2016}, indicating that the beam width could be narrower when performing optogenetic stimulation in \emph{in vivo} experiments.

\newpage
\putbib 

\end{bibunit}

\end{document}